\documentclass[iop]{emulateapj}
\begin{document}

\title{The 4.5~$\micron$ full-orbit phase curve of the hot Jupiter HD~209458b}

\author{Robert T. Zellem$^{1}$, Nikole K. Lewis$^{2,11}$, Heather A. Knutson$^{3}$, Caitlin A. Griffith$^{1}$, Adam P. Showman$^{1}$, Jonathan J. Fortney$^{4}$, Nicolas B. Cowan$^{5}$, Eric Agol$^{6}$, Adam Burrows$^{7}$, David Charbonneau$^{8}$, Drake Deming$^{9}$, Gregory Laughlin$^{4}$, Jonathan Langton$^{10}$}

\affil{$^{1}$ Lunar and Planetary Laboratory, University of Arizona, 1629 E. University Blvd., Tucson, AZ 85721, USA}
\affil{$^{2}$ Department of Earth, Atmospheric and Planetary Sciences, Massachusetts Institute of Technology, 77 Massachusetts Avenue, Cambridge, MA 02139, USA}
\affil{$^{3}$ Division of Geological and Planetary Sciences, MC 170-25 1200 E. California Blvd., Pasadena, CA 91125, USA}
\affil{$^{4}$ Department of Astronomy \& Astrophysics, University of California, Santa Cruz, 1156 High Street, Santa Cruz, CA 95064, USA}
\affil{$^{5}$ Department of Earth and Planetary Sciences, Northwestern University, Technological Institute, 2145 Sheridan Road, Evanston, IL 60208, USA}
\affil{$^{6}$ Astronomy Department, University of Washington, Physics-Astronomy Bldg., 3910 15th Ave NE, Seattle, WA 98195 USA}
\affil{$^{7}$ Department of Astrophysical Sciences, Princeton University, 4 Ivy Lane, Peyton Hall, Princeton University, Princeton, NJ 08544, USA}
\affil{$^{8}$ Harvard-Smithsonian Center for Astrophysics, 60 Garden Street MS-16, Cambridge, MA 02138, USA}
\affil{$^{9}$ Department of Astronomy, University of Maryland, College Park, MD 20742, USA }
\affil{$^{10}$ Physics Department, Principia College, 1 Maybeck Place, Elsah, Illinois 62028, USA}
\affil{$^{11}$ Sagan Fellow}

\email{rzellem@lpl.arizona.edu}

\begin{abstract}

The hot Jupiter HD~209458b is particularly amenable to detailed study as it is among the
brightest transiting exoplanet systems currently known (V-mag = 7.65; K-mag = 6.308) and has a large planet-to-star contrast ratio. HD~209458b is predicted to be in synchronous 
rotation about its host star with a hot spot that is shifted eastward of the substellar point by
superrotating equatorial winds.  Here we present 
the first full-orbit observations of HD~209458b, in which its 4.5~$\micron$ emission
was recorded with $Spitzer$/IRAC. 
Our study revises the previous 4.5~$\micron$ measurement of HD~209458b's secondary eclipse emission downward by $\sim$35\% to $0.1391\%^{+0.0072\%}_{-0.0069\%}$,
changing our interpretation of the properties of its dayside atmosphere. We find that the hot spot on the planet's dayside is shifted eastward of the substellar point by $40.9\degr\pm{6.0\degr}$, in agreement with circulation models predicting equatorial superrotation. HD~209458b's dayside (T$_{bright}$ = 1499 $\pm$ 15 K) and nightside (T$_{bright}$ = 972 $\pm$ 44 K) emission 
indicates a day-to-night brightness temperature contrast smaller than that observed for more highly irradiated exoplanets,
suggesting that the day-to-night temperature contrast may be partially a function of the incident stellar radiation. The observed phase curve shape deviates modestly from global circulation model predictions potentially due to disequilibrium chemistry or deficiencies in the current hot CH$_{4}$ line lists used in these models. Observations of the phase curve at additional wavelengths are needed in order to determine the possible presence and spatial extent of a dayside temperature inversion, as well as to improve our overall understanding of this planet's atmospheric circulation.

\end{abstract}

\section{Introduction}
Of the more than 1100 transiting exoplanets discovered to date, over 150 are gas giant planets known as ``hot Jupiters'' that have near-Jupiter masses (0.5M$_{Jupiter}$ $\le$ M $\le$ 5M$_{Jupiter}$ ) and that orbit very close to their host stars (semi-major axis $a$ $\le$ 0.1 AU) . These transiting exoplanets are predicted to be tidally locked \citep[e.g.,][]{macdonald64, peale74} so that one hemisphere always points towards its host star while the other is in perpetual night. The resulting day-to-night temperature contrast is predicted to drive fast $\sim$1 km/s winds, which transfer heat from the dayside to the nightside hemisphere thereby shifting the substellar hotspot and decreasing the day-to-night temperature contrast \citep[e.g.,][]{showmanguillot02, coopershowman05, coopershowman06, showman08a, showman08b, langtonlaughlin07, langtonlaughlin08, dobbsdixonlin08, showman09}.

One of the best methods to directly constrain the nature of the atmospheric circulation patterns on these planets is to continuously monitor their infrared (IR) emission to characterize its full-orbit phase curve. Such observations yield longitudinal disk variations, which can then be transformed into a longitudinal temperature profile \citep{cowanagol08} to measure the redistribution of heat. The \emph{Spitzer Space Telescope} is the only platform currently capable of making mid- to far-IR full-orbit observations due to its stability, continuous viewing capability, and access to longer wavelengths than the \emph{Hubble Space Telescope} \citep{harrington06, cowan07, knutson07, knutson12, cowan12, lewis13, maxted13}.  


In this paper we present new phase curve observations of the hot Jupiter HD~209458b, which has a particularly favorable planet-star radius ratio \citep[R$_{p}$/R$_{s}$$ = 0.12086 \pm 0.00010$;][]{torres08} and orbits a bright star (V-mag = 7.65; K-mag = 6.308). HD~209458 is a relatively quiet G-type star \citep[e.g.,][]{charbonneau00, knutson10}, which has been shown to vary by less than $\sim$0.002 mag at visible wavelengths over a period of 58 days \citep{rowe08}. This stability aids in the interpretation of its measured phase curve, as the star can be assumed to remain constant at the level of our measurements. Radial velocity and secondary eclipse observations indicate that this planet has a circular orbit \citep{henry00, mazeh00, deming05, laughlin05, winn05, wittenmyer05, kipping08, southworth08, torres08, crossfield12}, and is therefore likely to be in a synchronous rotation state.


HD~209458b's IR dayside emission has been measured during secondary eclipse, when the planet passes behind its host star, with two IRTF/SpeX K and L-band eclipses \citep{richardson03}, one \emph{Spitzer}/MIPS 24 $\micron$ eclipse \citep{deming05}, two \emph{Spitzer}/IRS 7--13.2 $\micron$ eclipses \citep{richardson07}, two \emph{Spitzer}/IRS 7.46--15.25 $\micron$ eclipses \citep{swain08}, one eclipse simultaneously observed at the four \emph{Spitzer}/IRAC passbands \citep{knutson08}, one \emph{Hubble}/NICMOS 1.5--2.5 $\micron$ eclipse \citep{swain09}, and three \emph{Spitzer}/MIPS eclipses \citep{crossfield12}. Multiple studies have used these emission data to constrain HD~209458b's dayside composition and thermal profile \citep[e.g.,][]{burrows07, fortney08, madhuseager09, line14}. The current consensus is that this planet's dayside emission spectrum is best-matched by models with a temperature inversion, possibly from TiO and VO absorption, two molecules typical of cooler stellar atmospheres \citep{burrows07, knutson08, fortney08}. However, TiO and VO have yet to be conclusively detected \citep{desert08}. Some groups have also argued that TiO would likely be lost to cold traps in the planet's interior and on the night side \citep{showman09, spiegel09, parmentier13}. 





Phase-curve measurements indicate a planet's longitudinal brightness temperature variations, allowing for the measurement of heat redistribution from the day to night side. HD~209458b, with an equivalent temperature of 1450~K (assuming a zero albedo) is comparatively cooler than most of the other hot Jupiters for which temperature contrasts have been measured: HAT-P-2b \citep{lewis13}, HD 149026b \citep{knutson09a}, HAT-P-7b \citep{borucki09}, WASP-18b \citep{maxted13}, and WASP-12b \citep{cowan12} \citep[see Fig. 1 in][] {perezbeckershowman13}. Thus measurements of HD~209458b's day-to-night temperature contrast enable further exploration of the theory that this contrast is driven by insolation \citep{perezbeckershowman13}. In addition, the phase curve determines whether HD~209458b, like other hot Jupiters \citep[e.g., HD~189733b;][]{knutson07, knutson09b, knutson12}, has a hot spot shifted eastward of the substellar point. Such a shift would indicate equatorial superrotation, as first predicted by the global circulation model (GCM) of \citet{showmanguillot02}.

Here we analyze post-cryogenic \emph{Spitzer Space Telescope} \citep{werner04} IRAC2 \citep{fazio04} full-orbit IR observations of HD~209458b. Our observations probe HD~209458b's dynamical processes by measuring its 4.5~$\micron$ dayside emission, location of its hot spot, and day-to-night temperature contrast. Previous secondary eclipse data were obtained in a non-standard observing mode in which $Spitzer$ cycled continuously between the four IRAC subarrays \citep{knutson08}. This mode enabled coverage of all four bands during a single eclipse event but with one-fifth the effective cadence of the now standard staring mode observations \citep[e.g.,][]{knutson12, lewis13}. Here we observe continuously in one IRAC band alone, resulting in a higher cadence. We therefore expect that our new observations will allow for a more precise estimate of the 4.5~$\micron$ eclipse depth. Our full-orbit data also samples HD~209458b's phase curve and measures for the first time its nightside emission. A previous 24~$\micron$ full-orbit phase curve was corrupted by instrumental sensitivity variations \citep{crossfield12}. From these data, we compare the hot spot offset, day-night flux differences, and secondary eclipse emissions derived here for HD~209458b with GCMs that predict these values.

\section{Observations}
The observations were taken between 2010 January 17 20:55:24.4 UT and 2010 January 21 21:28:42.0 UT, resulting in 96.55 hours of data. They begin shortly before secondary eclipse, continue through primary transit, and end shortly after secondary eclipse. To minimize readout time and thereby maximize image cadence, the subarray mode was used in which 32$\times$32 pixel images were stored as sets of 64 in a single FITS datacube with a single image header. By assuming uniform spacing in time, the mid-exposure time for each image is calculated from the header keywords MBJD\_OBS (start of the first image in each cube), AINTBEG (integration begin), and ANTIMMEEND (integration end). Effectively, the image spacing is 0.4 seconds. We report our timing measurements using BJD\_UTC.


\subsection{IRAC2 4.5~$\micron$ Photometry}
The sky background is estimated in each image by first masking out the target star with a 10-pixel radius. Three sigma outliers are then trimmed over three iterations to remove hot pixels and cosmic ray hits from the set of 64 measurements at that pixel position from a given data cube. The filtered background counts are fit with a Gaussian function. This count is then subtracted from each image.



We determine the $x$- and $y$-position of HD~209458 in each image using a flux-weighting methodology similar to a centroid that employs a five-pixel-radius circular aperture \citep[e.g.,][]{knutson08}. We find the latter method results in a more stable position estimate than the center of a two-dimensional Gaussian fit, and a correspondingly smaller scatter in our final photometry, in good agreement with our previous results for $Spitzer$ 4.5~$\micron$ phase curve observations of HD~189733b \citep{knutson12} and HAT-P-2 \citep{lewis13}. We first calculate our photometry using a fixed circular aperture with radii equal to 2.0, 2.1, 2.2, 2.3, 2.4, 2.5, 2.6, 2.7, 2.8, 2.9, 3.0, 3.5, 4.0, 4.5, and 5.0 pixels.

We also explore apertures that scale according to the noise pixel parameter $\tilde{\beta}$, which is defined in Section 2.2.2 (ÒIRAC Image QualityÓ) of the IRAC instrument handbook as:
\begin{eqnarray}
\tilde{\beta} = \frac{(\sum I_{i})^{2}}{\sum(I_{i}^{2})}
\end{eqnarray}
in which $I_{i}$ is the intensity in a given pixel $i$. The noise pixel parameter is equal to one over the sharpness parameter $S_{1}$, \citep{mullerbuffington74}, and proportional to the full-width half-maximum (FWHM) of the stellar point-spread function (PSF) squared \citep{mighell05}; for a more thorough discussion of the noise pixel parameter, please see Appendix A of \citet{lewis13}. We generate two additional aperture populations using the noise pixel parameter: $\sqrt{\tilde{\beta}}\times$[0.6, 0.7, 0.8, 0.85, 0.9, 0.95, 1.0, 1.05, 1.1, 1.15, 1.2] and $\sqrt{\tilde{\beta}}+[-0.8, -0.7, -0.6, -0.5, -0.4, -0.3, -0.2, -0.1,$ 0, 0.1, 0.2, 0.3, 0.4]. We find that a fixed aperture size of 2.7 pixels produces the smallest amount of scatter in the final residuals, which is also consistent with our conclusions for previous 4.5~$\micron$ phase curve observations \citep{knutson12, lewis13}.


Outliers in the measured flux, $x$- and $y$-positions of the target, or the noise pixel parameter $\tilde{\beta}$ are removed by first discarding six sigma outliers and then by using a moving median filter with a width of 10 points and discarding outliers greater than four sigma. After applying this filter we found that there were two points in our observations (at BJD\_UTC $\approx 2455214.9$ and $\approx 2455217.29$) in which the star displayed a sudden, sharp excursion in position typically lasting several minutes. These excursions are most likely the result of micrometoerite hits and we exclude them from our subsequent analysis. The two excluded segments of the light curve contain a total of 984 images, corresponding to 0.13\% of the total data set. This raw, filtered data is presented in Table ~\ref{table:data}.

\begin{deluxetable*}{lcccc}
\tabletypesize{\scriptsize}
\tablecaption{Photometric Data}
\tablewidth{0pt}
\tablehead{\colhead{JD (BJD\_UTC  $-$ 2455214)} & \multicolumn{2}{c}{Raw, Filtered Photometry} &\multicolumn{2}{c}{Final, Corrected Photometry} \\
&  Relative Flux & Uncertainty & Relative Flux & Uncertainty}
\startdata

0.413525& 1.007950& 0.008475& 1.002779& 0.003201\\
0.413530& 1.009589& 0.008482& 1.004326& 0.003201\\
0.413535& 1.009016& 0.008479& 1.002103& 0.003201\\
0.413539& 1.003937& 0.008458& 0.997562& 0.003201\\
0.413544& 1.011562& 0.008490& 1.004216& 0.003201\\
0.413548& 1.006882& 0.008470& 0.998892& 0.003201\\
0.413553& 1.005686& 0.008465& 0.998323& 0.003201\\
0.413557& 1.009775& 0.008482& 1.002209& 0.003201\\
0.413562& 1.006700& 0.008469& 0.999762& 0.003201\\
0.413566& 1.003509& 0.008456& 0.996224& 0.003201\\

\enddata
\tablenotetext{}{Sample of the filtered, raw data (see Fig. ~\ref{fig:rawlc}) and the final, corrected photometry (see Fig. ~\ref{fig:finallc}).}
\label{table:data}
\end{deluxetable*}

\subsection{Flux Ramp Correction}
Previous $Spitzer$/IRAC observations \citep[e.g.,][]{beerer11, todorov12, knutson12, lewis13} note a ramp-like change in the observed flux at 4.5~$\micron$, lasting approximately one hour.  \citet{agol10} examined a similar effect in the 8 $\micron$ bandpass and postulated that the effect might be due to a combination of thermal settling of the telescope and charge-trapping in the array. This effect is mitigated by trimming the first hour of data after the start of the observation and again after the downlink break (at phase $\approx$ 0.75). We then fit the trimmed data with the ramp function given by \citet{agol10}, as part of our global fit described in Section \ref{sec:decorrelation}:
\begin{eqnarray}
F'/F = 1 + a_{1}e^{-t/a_{2}} + a_{3}e^{-t/a_{4}}
\end{eqnarray}
in which $F'$ is the flux with the ramp, $F$ is the flux corrected for the ramp, $a_{1}$--$a_{4}$ are the correction coefficients, and $t$ is the time. A Bayseian Information Criterion \citep[BIC\footnote{BIC = $\chi ^{2} + k \ln (n)$, in which $k$ is the number of free parameters and $n$ is the number of datapoints in the fit};][]{schwarz78} analysis indicates that a ramp correction is unnecessary in order to fit the trimmed data.  We therefore leave this ramp function out of our subsequent analysis, consistent with our previous 4.5~$\micron$ phase curve analyses \citep{knutson12, lewis13}.

\subsection{\emph{Spitzer}/IRAC Decorrelation} \label{sec:decorrelation}
\begin{figure}
\includegraphics[width=1\columnwidth]{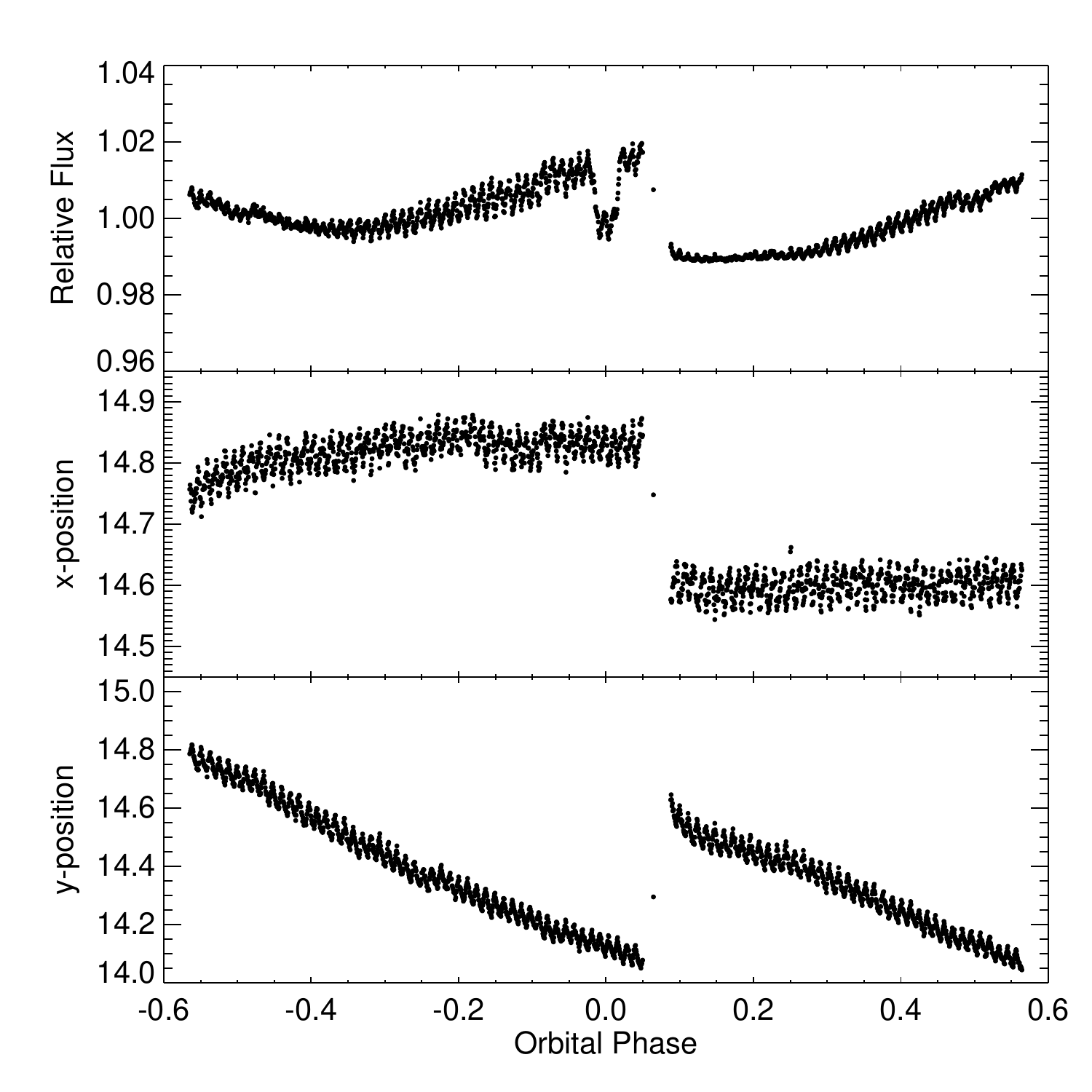}
\caption{Raw $Spitzer$/IRAC2 4.5~$\micron$ full-orbit photometry of the exoplanet HD~209458b (\emph{Top}), its $x$-position (\emph{Middle}), and $y$-position on the array (\emph{Bottom}) vs. the orbital phase after filtering. All frames are binned by 400 data points ($\approx$2.5 minutes). The gap in the data is due to spacecraft downlink. The raw flux is highly correlated with the $x$- and $y$-position on the array, which we remove here with a pixel-mapping Gaussian detrending method (see Section \ref{sec:decorrelation}).}
\label{fig:rawlc}
\end{figure}

Previous studies of transiting exoplanets find that the largest error source in $Spitzer$/IRAC 3.6 and 4.5~$\micron$ data is due to intra-pixel sensitivity variations, which cause variations in the telescope pointing to manifest as changes in the measured flux from the target star \citep[e.g.,][]{charbonneau08, beaulieu10, ballard10, knutson12, lewis13}. This effect can be easily seen in our 4.5~$\micron$ full-orbit data (Fig. \ref{fig:rawlc}) in which the raw flux tracks with the $x$- and $y$-positions on the detector. We de-trend the data using a pixel-mapping Gaussian weight decorrelation method \citep{ballard10, knutson12, lewis13} in which each photometric measurement is corrected by its 50 nearest neighbors in $x$-, $y$-, and $\tilde{\beta}$-space \citep[see Equation 2 in][]{knutson12}. We find, consistent with \citet{lewis13}, that inclusion of the noise pixel parameter $\tilde{\beta}$ terms increases the scatter in our final solutions. We therefore use the following simplified version of this correction instead:
\begin{eqnarray}\label{eqn:gw}
F_{meas,j} = F_{0,j} \sum_{i=0}^{n}e^{-(x_{i}-x_{j})^{2}/2\sigma_{x,j}^{2}} \times e^{-(y_{i}-y_{j})^{2}/2\sigma_{y,j}^{2}}
\label{eqn:gw_short}
\end{eqnarray}
in which $F_{meas,j}$ is the measured flux in the $j^{th}$ image; $F_{0,j}$ is the intrinsic flux; $x_{j}$ and $y_{j}$ are the measured $x$-position and $y$-position; and $\sigma_{x,j}$ and $\sigma_{y,j}$ are the standard deviations of the $x$ and $y$ vectors over the full range in $i$ (0 to $n$, where $n=50$ nearest neighbors). The 50 nearest neighbors to the $j^{th}$ image are determined via:
\begin{eqnarray}
d_{i,j} = \sqrt{(x_{i}-x_{j})^{2} + (y_{i}-y_{j})^{2}}
\end{eqnarray}
in which $d_{i,j}$ is the distance vector between the $j^{th}$ image and the $i^{th}$ image; $x_{i}$ and $x_{j}$ are the measured $x$-positions of the the $i^{th}$ and $j^{th}$ images, respectively; and $y_{i}$ and $y_{j}$ are the measured $y$-positions of the the $i^{th}$ and $j^{th}$ images, respectively. As a first-order estimation, we set the uncertainties on individual flux values equal to the standard deviation of the normalized residuals (SDNR) of the raw flux divided by the Gaussian weight (see Equation \ref{eqn:gw_short}).


We employ a Levenburg-Marquardt least-squares minimization via the Interactive Data Language (IDL) \texttt{mpfit} function \citep{markwardt09}  to solve simultaneously for the best-fit transit, secondary eclipse, and phase curve functions while optimizing the corresponding pixel map at each step in our minimization. We first use a \citet{mandelagol02} model light curve modified for full-orbit observations to calculate primary transit and secondary eclipse light curves using an initial guess for the system parameters, then divide those out of the raw light curve before making an initial pixel map. We next calculate a pixel map and corresponding correction (Equation \ref{eqn:gw}) for the intrapixel sensitivity variations at each point in the light curve. Thus, the Levenburg-Marquardt performs an iterative fit in which it finds the light curve model that, when the data is divided by the model, gives the best $x$- and $y$-position decorrelation (or pixel map) which minimizes the scatter in the corrected data/residuals.



For this analysis, the orbital period ($P$ = 3.5247455 days) and eccentricity ($e$ = 0) are fixed to values from a previous multi-transit study \citep{torres08} while the stellar limb darkening coefficients ($ld_{1-4}$ = [0.4614, -0.4277, 0.3362, -0.1074]) are derived from a three-dimensional model stellar atmosphere \citep{hayek12}. The inclination $i$, ratio of semi-major axis to stellar radius $a/R_{s}$, mid-transit time $T_{c}$, ratio of planetary to stellar radius $R_{p}/R_{s}$, secondary mid-eclipse times $T_{c}$, and secondary eclipse depths are left as free parameters. We also search for longitudinal brightness variations across the face of HD~209458b by fitting the full-orbit light curve with the \citet{cowanagol08} phase curve function:
\begin{eqnarray}
F = 1 + c_{0} + c_{1} \cos(2 \pi t/P) + c_{2} \sin(2 \pi t/P) \nonumber \\
+\,c_{3} \cos(4 \pi t/P ) + c_{4} \sin(4 \pi t/P )
\end{eqnarray}
in which $c_{0}$ is the secondary emission of the planet (here, we use the deeper measured secondary eclipse depth to establish this baseline so that it has a relative flux of unity), $c_{1}-c_{4}$ are free parameters, $t$ is the time, and $P$ is the planetary orbital period. The BIC indicates that inclusion of this phase curve function is necessary and that the $c_{3}$ and $c_{4}$ terms do not improve the quality of the fit; therefore only the $c_{1}$ and $c_{2}$ terms are used in the final fit. Fitting simultaneously for the phase curve, transit, and secondary eclipse models allows us to accurately account for the effect that the phase curve shape has on our estimates of the various transit and secondary eclipse parameters \citep{kippingtinetti10}. After finding an initial full-orbit solution to the dataset, the photometric uncertainties are then inflated by a factor of 2.2 in order to produce a global best-fit solution with a reduced $\chi^{2}$ equal to unity.


Next we explore the global solution-space with a Markov Chain Monte Carlo analysis \citep[MCMC; e.g.,][]{ford05}. The initial best-fit parameter solutions are randomly perturbed to seed six independent MCMC chains each consisting of $10^{5}$ links (steps). In each MCMC chain, a new trial solution is drawn from a Gaussian-distributed parameter space based upon the Levenburg-Marquardt best-fit parameters and associated uncertainties. Multiple (six) sufficiently long ($10^{5}$ links) MCMC chains are run with different randomly selected starting values to ensure both that an adequate amount of parameter space is sampled and that all chains converge to the global, and not a local, solution. After running each MCMC chain, we search for the point in which the chain has become well-mixed, as defined by where the $\chi^{2}$ value first falls below the median of all the $\chi^{2}$ values in the chain, and discard all links up to that ``burn-in'' point.  A Gelman-Rubin test \citep{gelmanrubin92} indicates a potential scale reduction factor $\le$ 1.07 for all parameters \citep{ford05}, suggesting that all six MCMC chains have converged to the same global solution.


Because MCMCs assume that the noise in the data is Gaussian and uncorrelated from one measurement to the next, they will typically underestimate the true uncertainties for data with a significant component of time-correlated noise \citep{carterwinn09}. To account for this ``red'' correlated noise, we independently estimate the uncertainties on our fit parameters using the ``residual permutation'' or ``prayer bead'' method \citep[e.g.,][]{jenkins02, southworth08, bean08, winn08}, in which the residuals from the best-fit MCMC solution are circularly permutated and added back onto the best-fit MCMC model to generate a new dataset. A Levenburg-Marquardt fit is applied to each of these new ``simulated'' data sets in order to build up histograms of the best parameter distributions from each permutation. We then compare the resulting uncertainties to their counterparts from the MCMC analysis and take the larger of the two for each fit parameter as our final uncertainties. The prayer bead uncertainties were up to 2.6 times larger than the corresponding MCMC uncertainties. Our final fit parameters and their associated one sigma uncertainties are reported in Table \ref{table:globalfitparams} and the final, corrected photometry is presented in Table \ref{table:data}.



\begin{figure}
\includegraphics[width=1\columnwidth]{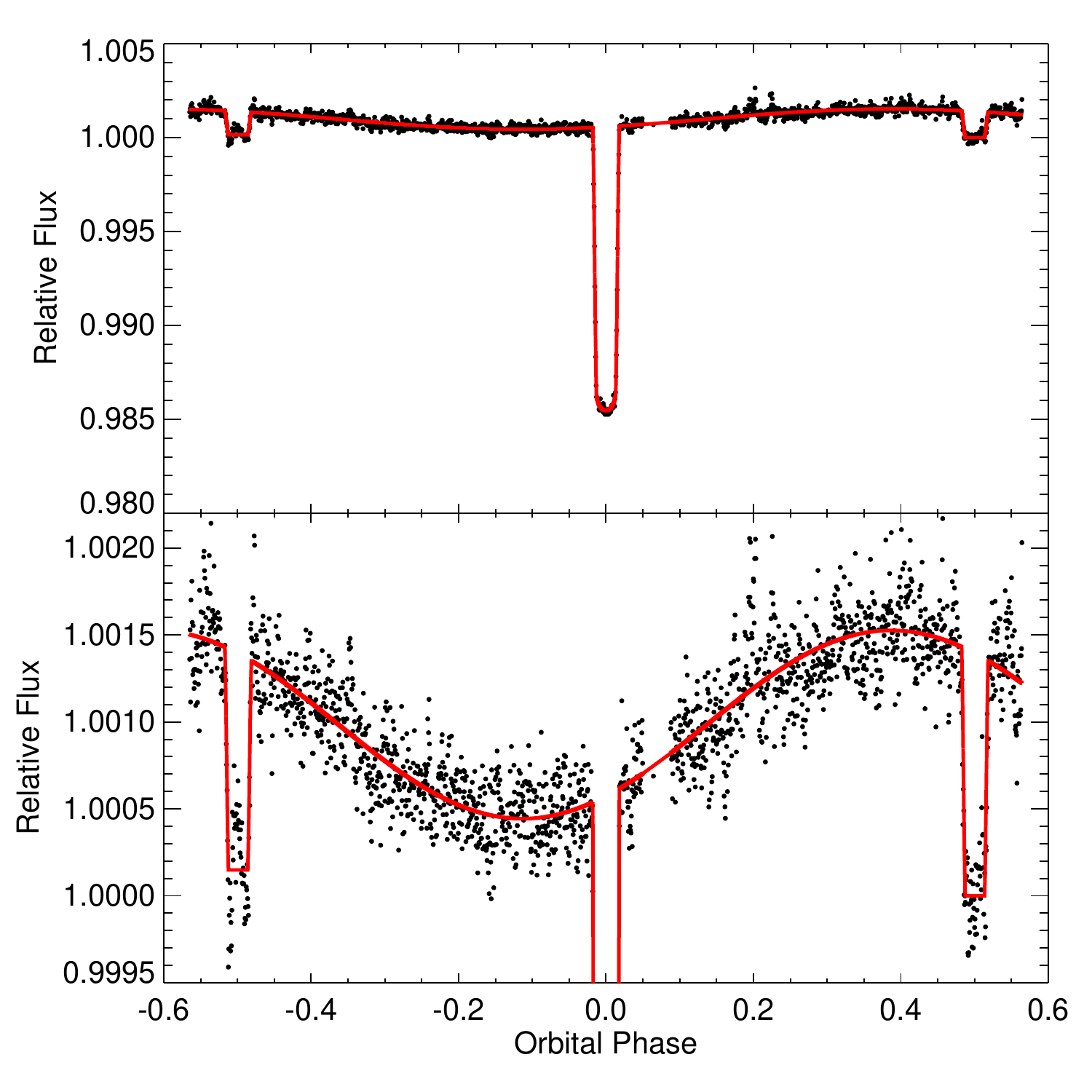}
\caption{Final, decorrelated full-phase 4.5~$\micron$ light curve of HD~209458b. The ``bumps'' just prior to the first second eclipse, in the middle of the first secondary eclipse, and at a phase $\approx$ 0.2 are likely due to residual systematic errors or possibly stellar activity (see Section \ref{sec:residualsystematics}).}
\label{fig:finallc}
\end{figure}

\begin{deluxetable*}{lcccc}
\tabletypesize{\scriptsize}
\tablecaption{Global Fit Parameters}
\tablewidth{0pt}
\tablehead{\colhead{Parameter} &  \colhead{Value} & \multicolumn{2}{c}{Uncertainty}}

\startdata

\emph{Orbital Parameters} &  &\\
\emph{i} (radians)   &      1.5131&${-    0.0017}$&${+    0.0016}$\\
\emph{a/R$_{s}$} &           8.810&${-     0.069}$&${+     0.064}$\\
&&\\
\emph{Primary Transit Parameters} && \\
\emph{T$_{c}$} (BJD$-2455000$) &         216.405640&${-   0.000097}$&${+   0.000091}$\\
\emph{R$_{p}$/R$_{s}$} &            0.12130&${-   0.00031}$&${+   0.00028}$\\
&&\\
\emph{Secondary Eclipse Parameters} && \\
{First Secondary Eclipse Depth} &       0.1243\%&${-   0.0067\%}$&${+   0.0073\%}$\\
\emph{T$_{c}$} (BJD$-2455000$) &              214.6462&${-    0.0012}$&${+    0.0011}$\\
{Second Secondary Eclipse Depth}    &     0.1391\%&${-   0.0069\%}$&${+   0.0072\%}$\\
\emph{T$_{c}$} (BJD$-2455000$) &              218.1694&${-    0.0011}$&${+    0.0010}$\\
{Average Secondary Eclipse Depth}    & 0.1317\%&${-   0.0048\%}$&${+   0.0051\%}$\\
&&\\
\emph{Phase Curve Parameters} && \\
\emph{c$_{1}$} &    $-0.0410\%$&${-   0.0046\%}$&${+   0.0051\%}$\\
\emph{c$_{2}$} &       0.0354\%&${-   0.0063\%}$&${+   0.0062\%}$\\
{Amplitude} &        0.109\%&${-   0.011\%}$&${+   0.012\%}$\\
{Minimum Flux} &          1.000443&${-   0.000067}$&${+   0.000068}$\\
{Minimum Flux Offset (hr)} &             $-9.6$&${-       1.4}$&${+       1.4}$\\
{Minimum Flux Offset ($\degr$)} &       $-40.9$&${-       6.0}$&${+       6.0}$\\
{Maximum Flux} &           1.001527&${-   0.000036}$&${+   0.000036}$\\
{Maximum Flux Offset (hr)}   &          $-9.6$&${-       1.4}$&${+       1.4}$\\
{Maximum Flux Offset ($\degr$)}   &    $-40.9$&${-6.0}$&${+6.0}$\\

\enddata
\label{table:globalfitparams}
\end{deluxetable*}

\begin{deluxetable}{lcc}
\tabletypesize{\scriptsize}
\tablecaption{Brightness Temperatures}
\tablewidth{0pt}
\tablehead{\colhead{Parameter} & & \colhead{Temperature (K)}}
\startdata

\emph{Secondary Eclipse Parameters} && \\
{First Secondary Eclipse} &&   1380 $\pm$ 32\\
{Second Secondary Eclipse}    && 1443 $\pm$ 30\\
{Average Secondary Eclipse} &&  1412 $\pm$ 22\\
&&\\
\emph{Phase Curve Parameters} && \\
{Amplitude} &&     527 $\pm$ 46\\
{Minimum Flux} &&   972 $\pm$ 44\\
{Maximum Flux} &&    1499 $\pm$ 15\\

\enddata
\label{table:brighttemps}
\end{deluxetable}

\subsection{Residual Systematics} \label{sec:residualsystematics}
Despite the success of this reduction, some systematics still remain. For example, small oscillations in flux occur before and during the first secondary eclipse, and again near an orbital phase of 0.2. These features are not correlated with any change in $x$, $y$, or $\tilde{\beta}$, suggesting that they are not associated with any changes in $Spitzer's$ pointing. In addition, they do not correspond with any background flux changes. These features could be due to stellar brightness variations from HD~209458, as their 0.1\% amplitude is similar to the amplitude of flux variations measured for the Sun \citep{eddy09}. However, the shape of the observed flux oscillations does not appear to match the sharp rise and slow decline expected for stellar flares \citep[e.g., Fig. 2 in][]{gary12}, and the time scale is too short for rotational spot modulations. Based on HD~209458's Ca II H and K emission measurements \citep[log$(R'_{HK})=-4.97$;][]{knutson10}, HD~209458 appears to have an activity level comparable to that of our own Sun \citep[log$(R'_{HK})=-4.96$;][]{noyes84}; therefore it might be reasonable to expect similar levels of infrared variability. \citet{knutson12} observed similar features in their $Spitzer$/IRAC HD~189733b data just after the first secondary eclipse at 3.6~$\micron$ and possibly before and after the second secondary eclipse at 4.5~$\micron$. This similarity suggests that the features in our light curve might instead be the result of residual uncorrected $Spitzer$/IRAC systematics. We leave a full exploration of these features for a future study and note that they do not significantly affect the overall shape of the phase curve, the primary transit, or the second secondary eclipse.


\begin{figure}
\includegraphics[width=1\columnwidth]{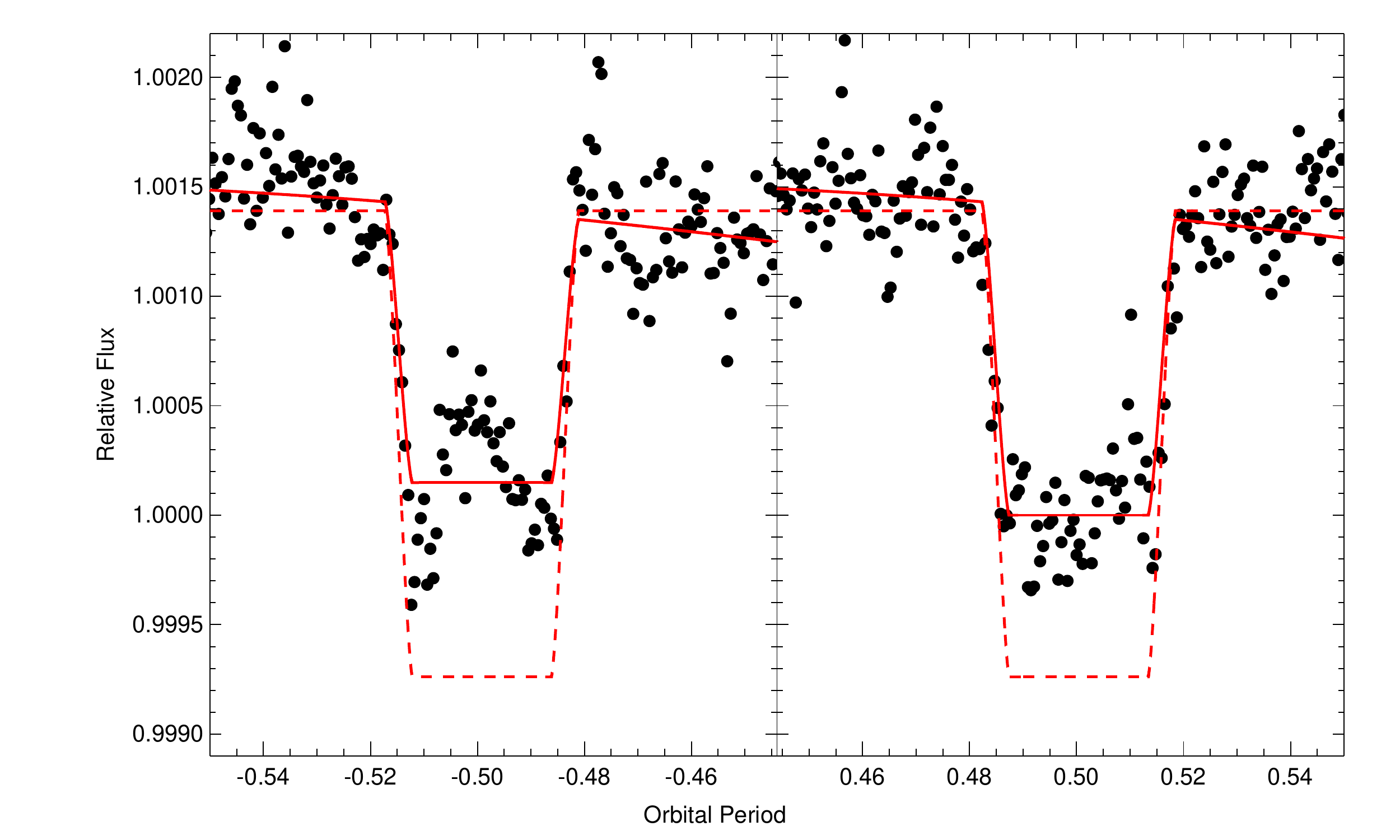}
\caption{Comparison of our two final, decorrelated 4.5~$\micron$ secondary eclipses of HD~209458b and corresponding best-fit model (red line) with the best-fit model of \cite{knutson08} (red dashed line). Here we find significantly ($\sim$35\,\%) shallower secondary eclipse values than \citet{knutson08}.}
\label{fig:finaleclipses}
\end{figure}


\section{Results}
\begin{figure}
\includegraphics[width=1\columnwidth]{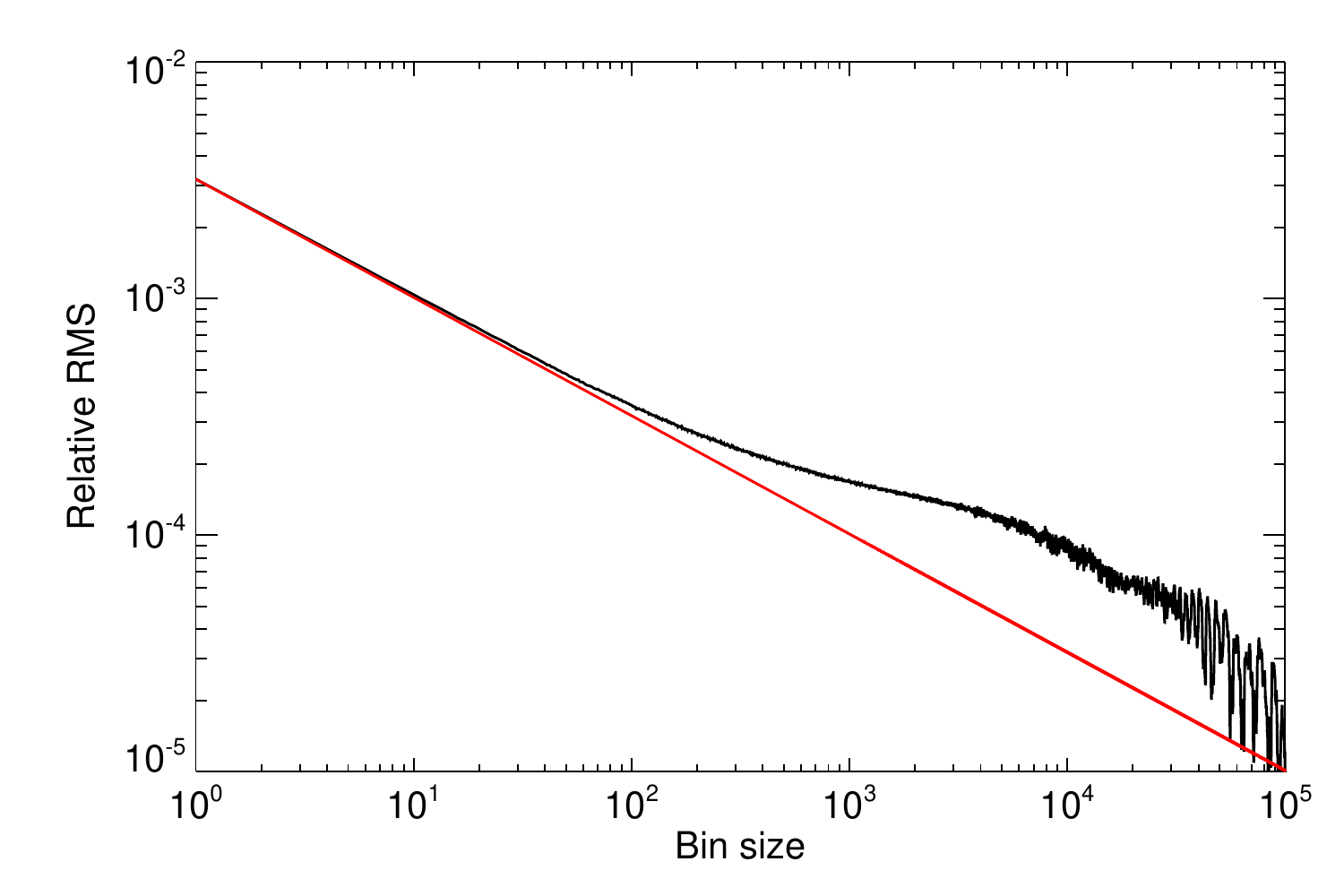}
\caption{Standard deviation of the residuals vs. bin size for the final, decorrelated 4.5~$\micron$ full-orbit data. The red line is assuming pure Gaussian noise ($\propto\frac{1}{\sqrt{N}}$, where $N$ is the bin size). Bin sizes of 10$^{2}$ and 10$^{3}$ correspond to time intervals of $\sim$0.7 minutes and $\sim$7.5 minutes, respectively.}
\label{fig:rel_RMS}
\end{figure}

\begin{figure}
\includegraphics[width=1\columnwidth]{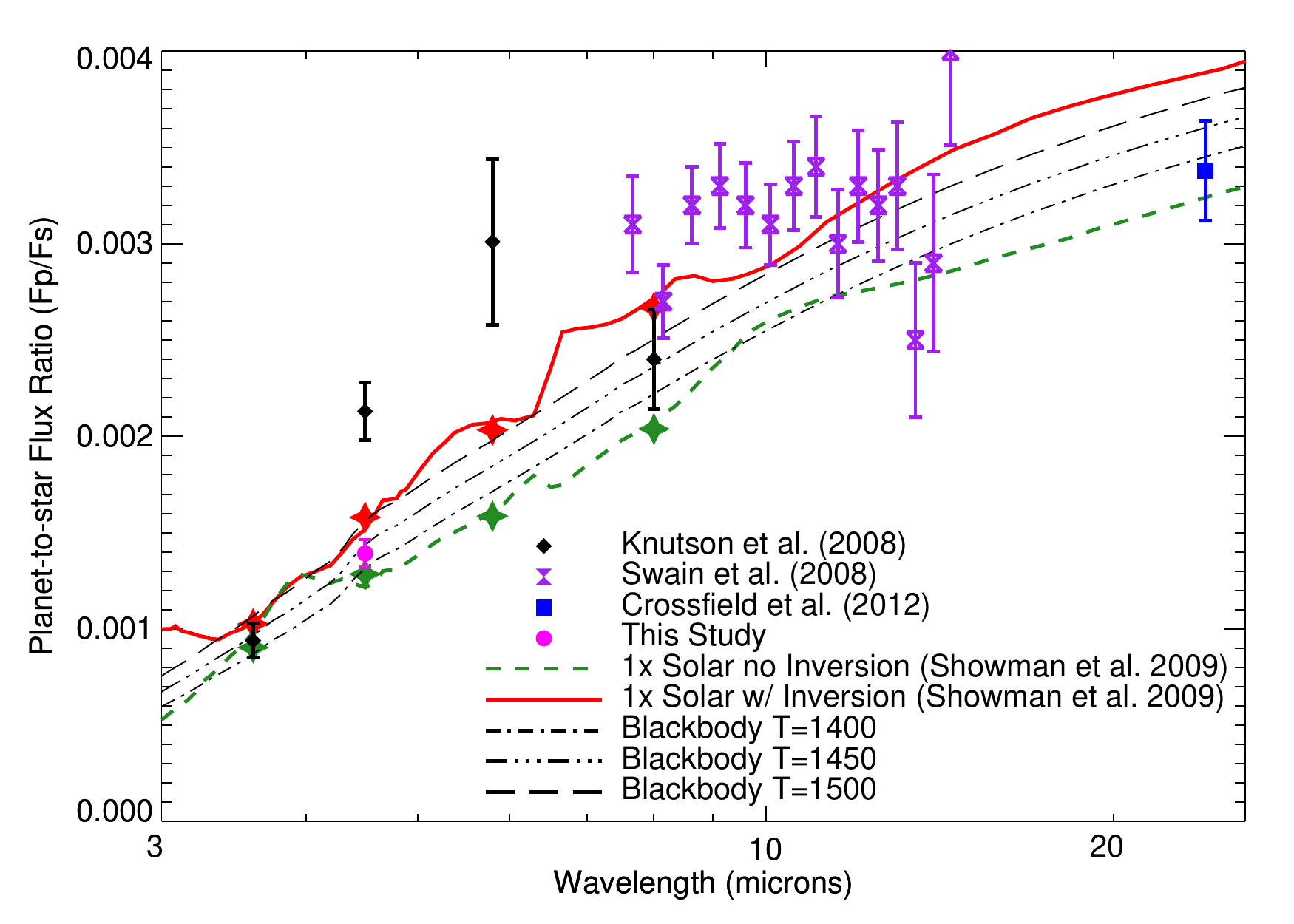}
\caption{Comparison of the planet-to-star flux ratio (F$_{p}$/F$_{s}$) emission measurements of HD~209458b by \citet{knutson08} (black diamonds), \citet{swain08} (purple hourglasses), \citet{crossfield12} (blue square), and this study (magenta circle; here, we plot the measurement from the second secondary eclipse $F_{p}/F_{s} = 0.1391\%_{-   0.0069\%}^{+   0.0072\%}$). Note how our revised 4.5~$\micron$ emission measurement brings this point into better agreement with the 1$\times$ solar abundance models with and without a thermal inversion (red line and green dashed line, respectively; these models are also binned to the IRAC bandpasses as indicated by the star symbol) by \citet{showman09}. Our new measurement also agrees reasonably well with blackbody emission curves with T=1400 (dash dot line), T=1450 (dash dot dot dot line), and T=1500 (long dash line).}
\label{fig:emission}
\end{figure}

The success of this reduction method is indicated by the decorrelated data (Fig. \ref{fig:finallc}) having a SDNR (the standard deviation of the normalized residuals) of 0.0032 which is within 14$\,\%$ (1.14 times) the photon noise limit. We estimate the significance of the red noise by calculating the RMS of the residuals in bins of increasing size, as shown in Figure \ref{fig:rel_RMS} and by using the equation from \citet{gillon06}:
\begin{eqnarray}\label{eqn:totnoise}
\sigma_{N}^{2} = \frac{\sigma_{w}^{2}}{N} + \sigma_{r}^{2}
\end{eqnarray}
where $\sigma_{N}$ is the measured uncertainty for a bin size $N$, $\sigma_{w}$ is the white noise and assumed to be equal to the SDNR, and $\sigma_{r}$ is the red noise. Using Equation \ref{eqn:totnoise}, we estimate the contribution of the red noise on relevant timescales. This value is 2.7\% of the total scatter in the relative flux corresponding to a 2.7 increase in the total noise in the 1 hour bins, the timescales of eclipse ingress and egress. Our final global solutions and their 1$\sigma$ uncertainties are listed in Table \ref{table:globalfitparams}. The best-fit planet-star radius ratio is $R_{p}/R_{s} =  0.12130_{-   0.00031}^{+   0.00028}$, which is in good agreement with the previously published 4.5~$\micron$ measurement of 0.12174 $\pm$ 0.00056 \citep{beaulieu10}. We find a secondary eclipse depth ($0.1391\%_{-   0.0069\%}^{+   0.0072\%}$) that is inconsistent with the previously published value from \citet{knutson08} at the 4.4 sigma level (see Figs. \ref{fig:finaleclipses} and \ref{fig:emission}); we discuss possible reasons for this disagreement in Section \ref{sec:secondaryeclipsediscussion}.



\section{Discussion}

\subsection{Secondary Eclipse Emission}\label{sec:secondaryeclipsediscussion}
The revision of the previous 4.5~$\micron$ emission measurement by \citet{knutson08} is significant because the high brightness temperatures at
4.5 and 5.8 $\micron$ were interpreted as evidence for a thermal inversion \citep{burrows07, knutson08, madhuseager09, madhuseager10, line14}. Our revision of HD~209458b's  4.5~$\micron$ emission is not without precedent. The original studies of both HD 189733b \citep{charbonneau08} and HD~209458b \citep{knutson08} were carried out by continuously cycling between the four IRAC arrays in order to cover all four bands during a single secondary eclipse. This method reduced the effective cadence of these observations to one-fifth that of the now standard staring mode observations, in which the star is observed continuously in a single band. The decision to cycle between detectors resulted in a significantly higher level of pointing jitter and correspondingly large flux variations in the resulting light curves as compared to staring mode observations. The disadvantages of this mode became apparent soon after these data were taken, and no other stars were observed in this manner. Although the original data were reduced using the standard techniques available at the time (namely fitting with a polynomial function of $x$- and $y$-position), it would not be surprising if the large pointing jitter resulted in a biased estimate of the eclipse depth. Subsequent observations of HD 189733b by \citet{knutson12} used the now standard staring mode and pixel mapping techniques, and found a 3.6~$\micron$ eclipse depth that deviated by approximately 7.5 sigma from the previous value. It is therefore not surprising that our new staring mode observations of HD~209458b also result in a 4.4 sigma revision to the published eclipse depth.

Our revised estimate for the 4.5~$\micron$ brightness temperature agrees significantly better with published models
for this planet (Fig. \ref{fig:emission}). Despite the original interpretation of the  \citet{knutson08} data as resulting from a
thermal inversion, it is worth emphasizing that even one-dimensional (1D) radiative-equilibrium models with several
tunable free parameters \citep{burrows07, fortney08} had difficulty simultaneously explaining all
four IRAC observations. Models in which the relative
abundances and pressure-temperature profiles were allowed to vary as free parameters were able to match all
four of the original measurements \citep{madhuseager09, madhuseager10, line14}.  These ``data-driven'' best fit models find the temperature inversion placed where they have the most leverage over the 4.5 and 5.8 $\micron$ bandpasses.  GCM simulations including a hot stratosphere generally do not produce elevated brightness temperatures at 4.5 and 5.8~$\micron$ relative to the other IRAC bandpasses \citet{showman09}, since the pressures probed by the IRAC bandpasses largely overlap, and the depth of the inversion is not a free parameter, but is controlled by the optical and infrared opacity sources. As shown in \citet{showman09}, in models where pressure-temperature profiles are not left as a free variable, much of the flux in the IRAC bandpasses is emitted to space from the bottom edge of, or even below, the
inversion, implying that high stratospheric temperatures
do not necessarily exert significant leverage on the flux emitted at
4.5 and 5.8 $\micron$ in these models.  Furthermore, the inversion covers only part
of the dayside in the 3D models, which also mutes its effect. However, there could be an absorber not accounted for in the models that gives rise to an inversion at a different level.


Our new analysis is in better agreement with the predictions of the general circulation models for this planet from \citet{showman09}, including cases both with and without dayside temperature inversions, and also blackbody models (see Fig. \ref{fig:emission}).  Although our new measurement does not rule out a dayside temperature inversion, it suggests that an inversion may not be necessary to explain the current 4.5~$\micron$ broadband data. New high-precision eclipse measurements in the other $Spitzer$ bandpasses, specifically photometry at 5.8 and 8~$\micron$, and ideally mid-infrared spectroscopic measurements \citep[e.g.,][]{madhuseager10, burrows14, line14} would help to provide a more definitive answer to this question.

\subsection{Phase Curve}
Our derived 4.5~$\micron$ light curve indicates an eastward shifted
hot spot, similar to the one observed for HD~189733b \citep{knutson07, knutson09b, knutson12}. Such eastward offsets were first predicted for hot Jupiters by
\citet{showmanguillot02},  and subsequent 3D circulation models
confirm that they are a robust feature of the hot-Jupiter circulation
regime, at least over a certain range of incident stellar fluxes and atmospheric
radiative time constants \citep[e.g.,][]{coopershowman05, showman08, showman09, showman13, menourauscher09, rauschermenou10, rauschermenou12, heng11a, heng11b, perna12, dobbsdixon12}.
In these circulation models, the eastward offset of the hot spot results
from advection of the temperature field by a fast, broad eastward equatorial
jet stream --- so-called equatorial superrotation.  In turn, the
equatorial superrotation results from interactions with the mean flow of
standing, planetary-scale waves induced by the day-night heating
contrast \citep{showmanpolvani11}.  When the radiative time constant is
comparable to the characteristic timescale for air parcels to advect
eastward over a planetary radius, then significant eastward hot spot offset from the substellar point should occur.  Overall, our observations
are consistent with this body of theory and models, and suggest that HD~209458b likely
exhibits equatorial superrotation at photospheric levels.

\begin{figure}
\includegraphics[width=1\columnwidth]{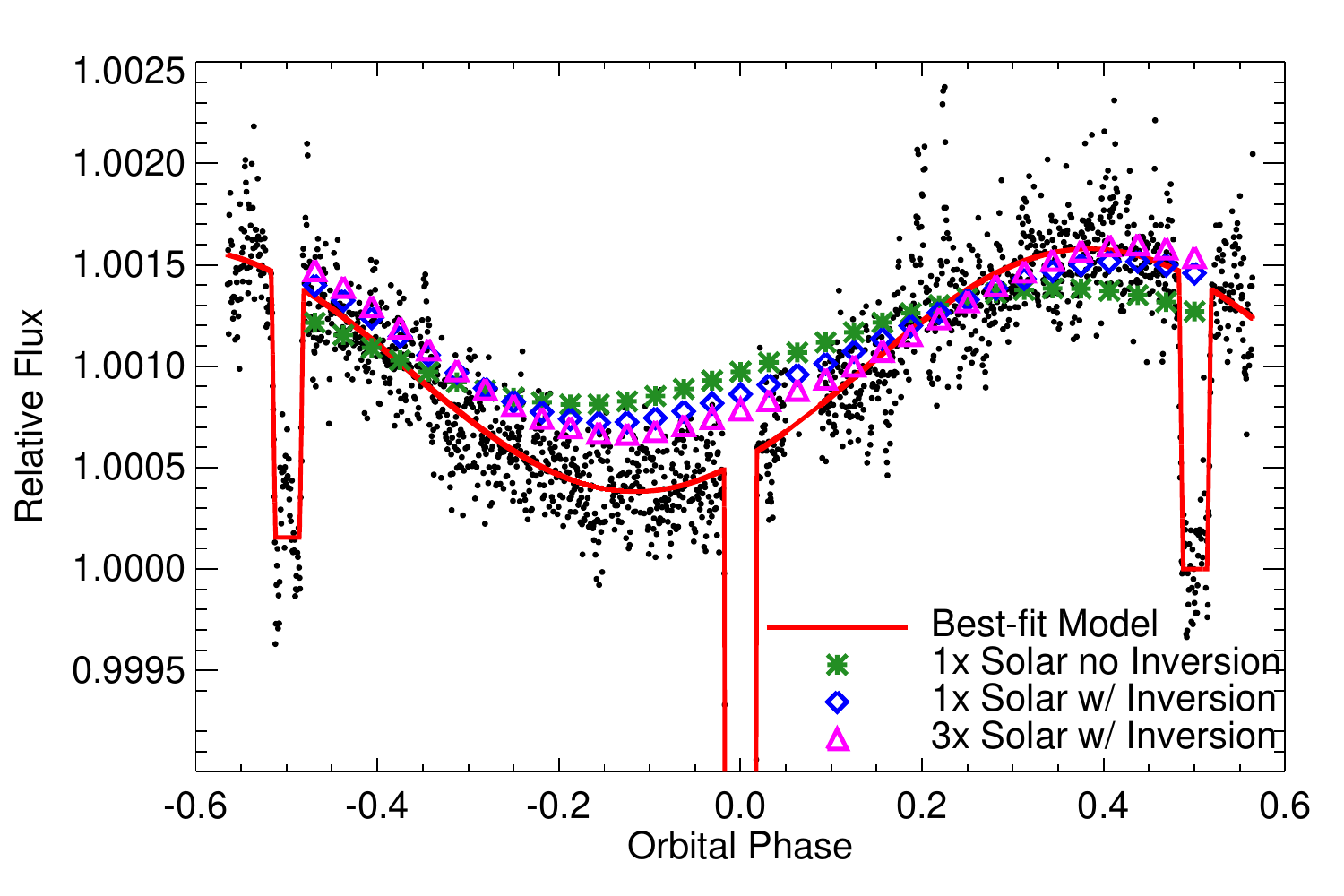}
\caption{Comparison of our reduced data (black points) and its best-fit phase curve model (red line) to the \citet{showman09} GCMs with 1$\times$ solar abundance and no thermal inversion (green astericks), an inversion with 1$\times$ solar abundance (blue diamonds), and an inversion with 3$\times$ solar abundance (magenta triangles). The \citet{showman09} inversion model predictions (blue diamonds and magenta triangles) nominally fit our observations, except on the nightside in which the GCM over-predicts HD~209458b's flux.}
\label{fig:sparccomp}
\end{figure}

Figure \ref{fig:sparccomp} compares our best-fit model of HD~209458b's observed full-orbit
flux  to theoretical phase curves generated from the global circulation models (GCMs) presented in \citet{showman09}. The \citet{showman09} 3D dynamical models of HD~209458b are the first with non-grey
radiative transfer across the entire wavelength range from the visible
through the infrared.  In our observations, we find a phase curve maximum that
occurs $9.6\pm{1.4}$ hours before secondary eclipse, corresponding
to a hot spot shifted $40.9\degr\pm{6.0\degr}$ eastward of the substellar point.
The theoretical phase curves generated from the GCM are shown for no-thermal-inversion models (i.e.,
models lacking TiO and VO), and models with TiO and VO included at 1$\times$ and 3$\times$
solar abundance. These molecules cause a thermal inversion due to the extreme opacity
of TiO and VO in the visible.\footnote{TiO and VO are simply proxies for any
visible-wavelength absorber in these models, and any other strong visible
absorber would lead to qualitatively similar behavior.} Interestingly, our
observations exhibit a peak flux in the light curve which leads secondary eclipse by
an amount intermediate to that predicted from the no-inversion and thermal-inversion
models.   Overall, our observed dayside emission (phase $\sim$0.25 to 0.75
in Figure \ref{fig:sparccomp}) matches the predictions of the \citet{showman09} models well,
especially for models including visible absorbers and a dayside inversion.

There exist a variety of factors that can influence the offsets of
the hot spot for a given hot Jupiter.
Generally speaking, a larger atmospheric opacity would move the
photospheres to higher altitude (lower pressure), in which the radiative time constant
is shorter, leading to a smaller hot-spot offset; and conversely for
smaller atmospheric opacity \citep{dobbsdixonlin08, showman09, lewis10, heng11b}.  Such opacity
variations could result from the gas metallicity (higher metallicity
generally implies greater opacity and vice versa), or potentially
from hazes, though there is currently no strong evidence that
hazes significantly affect the emission on HD~209458b \citep{deming13}. Moreover, for
specified opacities, the existence of atmospheric frictional drag could
lead to slower wind speeds, decreasing the hot spot offset, potentially even
to zero if the drag is sufficiently strong \citep{rauschermenou12, showman13, dobbsdixon12}.  Such drag could
result from Lorentz forces due to the partial thermal ionization
at high temperatures \citep{perna10, rauschermenou13, rogersshowman14}. Nevertheless, the fact that our observed
offset is significantly nonzero (with 6.9 sigma confidence) and agrees reasonably well with
GCM simulations performed in the absence of strong drag at
photospheric levels (Fig. \ref{fig:sparccomp}; \citet{showman09}) suggests that
these magnetohydrodynamic (MHD) effects do not play a dominant
role in controlling the hot spot offset for HD~209458b.



HD~209458b's observed phase curve, which represent the first 4.5~$\micron$ measurements of its nightside emission, indicates that HD~209458b has a day-to-night temperature contrast of $\Delta$T$_{obs} = 527 \pm 46$ K. We find that HD~209458b has a smaller contrast at 4.5~$\micron$  (A$_{obs} = 0.352 \pm 0.031$ \footnote{$A_{obs} = (flux_{day} - flux_{night})/flux_{day}$ \citep{perezbeckershowman13}}) than other hot Jupiters with higher levels of incident flux \citep[e.g., WASP-12b and WASP-18b;][respectively]{cowan12, maxted13}, consistent with the idea that this temperature contrast is driven by insolation \citep{perezbeckershowman13} as the radiative time constant decreases with increasing temperature \citep{showmanguillot02}. However, the only day-night temperature contrast measurements that exist for HD~209458b are at 4.5~$\micron$ (presented here) and an upper-limit at 8 $\micron$ \citep{cowan07}. Thus full-orbit phase curve measurements both at additional wavelengths, such as 3.6~$\micron$, and for other targets are necessary to confirm the hypothesis of \citet{perezbeckershowman13}.

In addition, HD~209458b's phase curve suggests that the nightside is much cooler than predicted by \citet{showman09}. Yet, while the location of the 1$\times$ and 3$\times$ solar abundance inversion phase curve minima \citep{showman09} more closely agree with our observations compared to the no inversion model, the GCM overestimates the night side flux, thereby underestimating the total day-to-night temperature contrast ($\Delta$T$_{GCM}$ $\approx$ 300 K vs. $\Delta$T$_{obs} = 527 \pm 46$ K). The \citet{showman09} models similarly overpredict HD~189733b's nightside emission compared to $Spitzer$/IRAC full phase observations by \citet{knutson12}. This discrepancy is attributed to disequilibrium carbon chemistry not included in the GCM, in particular quenching, which would increase the abundances of CO and CH$_{4}$ at higher altitudes so that measurements would be probing a comparably higher optically thick layer with cooler temperatures. In the HD~209458b abundance profiles of \citet{showman09} and \citet{moses11}, the vertical CO abundance profile is driven by chemical equilibrium, resulting in a relatively constant mixing ratio from 10 to 10$^{-8}$ bars. Therefore vertical quenching of CO would have a minimal effect on its abundance profile, suggesting that vertical quenching of CO is likely not the culprit for HD~209458b's cooler observed nightside. However, the \citet{moses11} abundance profiles of CH$_{4}$ suggest that vertical quenching can increase the CH$_{4}$ abundance by nearly an order of magnitude from 1 to 10$^{-5}$ bars compared to the equilibrium-driven \citet{showman09} profiles. This additional CH$_{4}$ could potentially help radiate heat and result in an overall cooler nightside than predicted by the GCM. To determine if quenched CH$_{4}$ is causing the nightside cooling, we are in the process of measuring HD~209458b's full-orbit phase curve at 3.6~$\micron$, which overlaps the CH$_{4}$ $\nu_{3}$ band. Analyses of these data with future GCM studies, which include both non-equilibrium chemistry and new hot CH$_{4}$ line lists \citep{exomol14}, will better constrain the properties of these two carbon-bearing species and the nightside cooling mechanism.

\section{Conclusions}
Here we present the first measurements of HD~209458b's 4.5~$\micron$ full-orbit phase curve using $Spitzer$/IRAC. Our data indicate 0.69 sigma agreement with a previous primary transit depth by \citet{beaulieu10} and revise HD~209458b's secondary eclipse emission measurement by \citet{knutson10} downward by $\sim$35$\,\%$, potentially weakening the evidence for a dayside temperature inversion. The phase-curve observations suggest both a hot spot shifted eastward of the substellar point and a day-to-night temperature contrast smaller than that of more highly irradiated hot Jupiters, suggesting that this contrast may be driven by the incident stellar flux. The shape of the phase curve, specifically the location and brightness temperature of the hot spot, suggests that HD~209458b could have a dayside inversion at a pressure level that is between that predicted by non-inversion models and that predicted by TiO and VO induced thermal inversion models ($\ga$0.008 bar) \citep{showman09}. However, new GCMs that include non-equilibrium chemistry and hot CH$_{4}$ lines \citep{exomol14} are necessary not only to confirm this hypothesis but also determine why HD~209458b's nightside is cooler than previously predicted. Thus, while we cannot draw any strong conclusions on HD~209458b's thermal inversion, we hope to better constrain the existence of a thermal inversion with upcoming full-orbit phase curve observations at 3.6~$\micron$ and eclipse mapping at both 3.6 and 4.5~$\micron$.

\section{Acknowledgements}
RZ and CAG are supported by the NASA Planetary Atmospheres Program.

NKL performed this work in part under contract with the California
Institute of Technology (Caltech) funded by NASA through the Sagan
Fellowship Program executed by the NASA Exoplanet Science Institute.

APS is supported by the NASA Origins program.

RZ would like to thank Travis Barman, Ian J. M. Crossfield, Julien de Wit, Davin Flateau, Joe Giacalone, Tiffany Kataria, Michael R. Line, Julianne I. Moses, Kyle A. Pearson, Emily Rauscher, Tamara M. Rogers, David K. Sing, and Mark R. Swain for their helpful discussions.

We would like to thank the two referees for their helpful comments and suggestions.

\bibliography{references}

\end{document}